\documentclass[aps,showpacs,twocolumn,superscriptaddress]{revtex4}
\usepackage{epsfig}
\usepackage{graphicx}
\usepackage{color}

\begin{document}

\title{Higher Order Harmonics of Modulational Instability Induced by
Vacuum Fluctuations}
\author{David Amans}
%\email{damans@ulb.ac.be}
\affiliation{Institut de Micro\'{e}lectronique,
Electromagn\'{e}tisme et Photonique, ENSERG,
23 av. des martyrs,
B.P. 257,
38016 Grenoble,
France
}
\author{Edouard Brainis}
%\email{ebrainis@pop.ulb.ac.be}
\affiliation{Optique et Acoustique,
CP 194/5, Universit\'{e} Libre de Bruxelles, Avenue F. D. Roosevelt
50, 1050 Bruxelles, Belgium}
\author{Serge Massar}
%\email{smassar@ulb.ac.be}
\affiliation{Laboratoire d'Information
  Quantique and QUIC,
Universit\'{e} Libre de Bruxelles, CP 165/59, Av. F. D. Roosevelt 50, 1050
Bruxelles, Belgium}

%\date{\today}

\begin{abstract}
We study the higher order harmonics of scalar modulational
instability in the regime where it arises spontaneously through
amplification of vacuum fluctuations. We obtain detailed
predictions concerning the detunings, intensities, growth rates and spectral widths
of the harmonics. These predictions are well verified by experimental results obtained by propagating high intensity light pulses through  optical
fibers.
\end{abstract}


\maketitle

\section{Introduction}

In a non linear dispersive medium the propagation of a continuous
wave may be impeded by the phenomena of Modulational Instability
(MI) whereby the continuous wave breaks up into a train of
localized pulses. This effect has been predicted and
verified in many branches
of physics, including fluid dynamics\cite{fd}, plasma
physics\cite{pp}, non linear optics\cite{nlop,HB,Tai86a,Agrawal95},
Bose-Einstein condensates\cite{bec}.

In the spectral domain the instability
induces the appearance of new frequencies whose amplitude grows
exponentially.
As the dynamical process continues more and more energy is converted from the monochromatic wave to the sidebands. Simultaneously harmonics of the unstable frequencies appear and also grow exponentially.

Although the process of MI is by now textbook material, the subsequent dynamics, and in  particular the appearance of the higher harmonics has been relatively little studied.
The main work we are aware about in this context are theoretical \cite{AK} and experimental  \cite{VSEH} investigations in the context of the Fermi-Paska-Ulam recurrence.
It is important to note that in these works the initial noise was taken
to be a classical wave at a specific frequency and the analysis therefore
involves a discrete set of modes.

In the present work we study both theoretically and experimentally
the growth of the higher order harmonics of MI in the regime where
the instability arises by spontaneous amplification of vacuum
fluctuations. The essential difference with respect to the earlier
work mentioned above is that in the present case a continuous
spectrum of vacuum fluctuations is involved (rather than a discrete
set of modes). We will characterize each harmonic by its detuning,
its intensity and growth rate, and its spectral width. Several of
these features would be different, or would simply not appear (in
the case of the spectral width), if one was dealing with a discrete
set of modes.

Our investigation is carried out in the regime of large gain, when many photons have been created in the sidebands, but before we reach the regime of pump depeletion (ie. before the energy in the side bands becomes comparable to the energy in the 
initial the monochromatic wave). One of the main interests of our work is to show how, even though we are working in a semiclassical regime where many photons are produced, the fact that the instability is seeded by vacuum flucutations leaves a distinct signature in the spectrum of the sidebands.

Our analytical predictions are well reproduced in our experimental
demonstration, based on light propagation through optical fibers.
This system is particularly well suited for the such investigations
because light propagation in low loss silica fibers is accurately
described by the Non Linear Schr\"odinger Equation (NLSE), hence
suitable to precise theoretical modelling, and because it is fairly
easy to work in a regime where initial noise is dominated by vacuum
fluctuations. In fact the appearance of the first harmonic -probably
seeded by vacuum fluctuations- was already reported in the first
experimental investigation of MI in optical fibers\cite{Tai86a}.
Thus our approach provides quantitative explanation for the complex
spectra -readily observable in experiments- that arise in MI seeded
by vacuum fluctuations in the regime of large gain. A related
comparison, but between numerical simulations and experimental
results- can be found in \cite{amansXP}.

Finally we note that the ideas and methods developed here are not
restricted to the problem of scalar MI, but should apply
{\em mutatis mutandis} to other forms of MI such as vectorial
MI, and more generally to any unstable dynamical system involving
a continuous set of modes (i.e. described by partial differential
equations), when the instability is seeded by vacuum fluctuations. We also note that our analysis can also be applied if the initial noise is classical white noise.

Section \ref{T1} contains our theoretical analysis. Our key result is eq. (\ref{MH1}) which encapsulates our predictions concerning intensities, growth
rates, frequencies, spectral widths of the harmonics. A discussion of the predictions provided by eq. (\ref{MH1}), of the hypotheses that go into its derivation, and of its interpretation, are given in section \ref{T2}.
Some background material, and details of some calculations have been relegated to Appendices. Section \ref{Expt} describes our experimental setup and shows that the experimental results are in good agreement with the theoretical predictions.

\section{Theoretical Analysis}\label{Theory}

\subsection{Harmonics of Modulational Instability}\label{T1}

 Let $A(x,\tau)$ be the slowly varying
field envelope of a light pulse propagating in
an isotropic single mode fiber with carrier frequency $\omega_0$,
where $x$ is the coordinate along the fiber and $\tau = t - x/v_g$
is the time variable of coordinates moving at group velocity
$v_g$. It obeys the NLSE (see Ref. \cite{Agrawal95})
\begin{equation}
i \partial_x A = \frac{\beta_2}{2}\partial_\tau^2 A- \gamma|A|^2A
\label{NLSE}
\end{equation}
where $\beta_2$ is the group velocity dispersion at frequency
$\omega_0$ and $\gamma = \frac{2
n_2 \omega_0}{
  \epsilon_0 c^2 n_0 A_{eff}}$ with $n_0$ the effective linear
refraction index for light guided in the fiber, $n_2$ the non
linear refraction index, $A_{eff}$ the effective cross section
of the fiber. With this normalization $ |A(x,\tau)|^2$ is the
instantaneous power flowing through the fiber at position $x$ at
time $\tau$.

We take the unperturbed solution to be
\begin{equation}
A=A_0 e^{i\phi_{NL}}
\label{A0}
\end{equation}
 with
$A_0$ a constant and $\phi_{NL}= \gamma A_0^2 x$ is the non-linear phase. As is well known, when $\beta_2 <0$ (the anomalous dispersion regime), the continuous solution eq. (\ref{A0}) is unstable. We give the "standard" derivation of this instability (following \cite{Agrawal95}) in Appendix \ref{AppA}, where we also include the predictions of quantum theory (when the instability is seeded by vacuum fluctuations) and discuss the regime of large gain.

In the present work we use a slightly different approach introduced by Hasegawa and Brinkman \cite{HB}. This will allow us to derive
many of the properties of the harmonics of MI (which are inaccessible using the standard method of Appendix \ref{AppA}).

Our starting point is to
parameterize the solution as:
\begin{equation}
A(x,\tau) = A_0 e^{i \phi_{NL}}(1 +
\epsilon(x,\tau))^{1/2}e^{i\sigma(x,\tau)}
\label{E1}
\end{equation}
where $\epsilon$ and $\sigma$ are real variables. Upon insertion of
this ansatz into the NLSE one obtains the equations
\begin{eqnarray}
\frac{\partial_x \epsilon}{1+\epsilon}- \beta_2 \left(
\frac{\partial_{\tau}\sigma\ \partial_{\tau}\epsilon }{1
+\epsilon}
+ \partial_{\tau}^2\sigma\right)&=&0,\nonumber\\
 \frac{\beta_2}{2}\left(
\frac{(\partial_{\tau}\epsilon)^2}{4 (1 + \epsilon)^2} -
\frac{\partial_{\tau}^2\epsilon}{2 (1 + \epsilon)} +
(\partial_{\tau}\sigma)^2\right) &&\nonumber\\
-\partial_x\sigma + \gamma |A_0|^2 \epsilon &=& 0. \label{HBE1}
\end{eqnarray}
The key to the approach of Hasegawa and Brinkman is to keep the non linear dependence of $A$ in $\epsilon$ and $\sigma$ in eq. (\ref{E1}), but to linearise the eqs. (\ref{HBE1}). Obviously this does not provide an exact solution to the NLSE, but it provides an approximate solution which captures many of the features of an exact solution (illustrated for instance by the good agreement between these theoretical predictions and our experimental results).

Upon linearisation of eqs. (\ref{HBE1}) one obtains the following
simplified set of equations:
\begin{eqnarray}
\partial_x {\epsilon}- \beta_2 \partial_\tau^2\sigma
=0 \quad ; \quad
%&=&0,\nonumber\\
-\partial_x \sigma - \frac{\beta_2}{4}\partial_\tau^2\epsilon +
\gamma
  |A_0|^2 \epsilon =0\ .
%&=&0.
\label{HBE2}
\end{eqnarray}
The Fourier transforms of $\epsilon$ and $\sigma$ are:
\begin{eqnarray}
\epsilon(x,\tau) &=&\frac{1}{\sqrt{2\pi}} \int_0^\infty d \omega \
\epsilon (x, \omega) e^{-i \omega \tau} + \epsilon^* (x, \omega)
e^{+i \omega \tau};
\nonumber\\
\sigma(x,\tau) &=&\frac{1}{\sqrt{2\pi}} \int_0^\infty d \omega \
\sigma (x, \omega) e^{-i \omega \tau} + \sigma^* (x, \omega) e^{+i
\omega \tau}. \nonumber\\
\label{AA1}
\end{eqnarray}
When $\beta_2 <0$ and $0<\omega^2<4 \gamma A_0^2 / |\beta_2|$ these
equations possess exponentially growing solutions:
\begin{eqnarray}
\epsilon(x,\omega) &=& \epsilon_+(\omega)e^{g x} +
\epsilon_-(\omega)e^{-g x}\ ,
\nonumber\\
\sigma(x,\omega)&=& \frac{g}{\omega^2 |\beta_2|}
\epsilon_+(\omega)e^{g x} - \frac{g}{\omega^2 |\beta_2|}
 \epsilon_-(\omega)e^{-g x}\ ,
\label{AA2}\end{eqnarray}
where
\begin{equation}
g  = \frac{|\beta_2| \omega}{2}\left( \frac{4 \gamma
  A_0^2}{|\beta_2|} - \omega^2 \right)^{1/2}.
\end{equation}

As discussed in Appendix \ref{AppA} one can also write the solution
to the NLSE as
$$
A(x,\tau) =  e^{i \phi_{NL}} \left(  A_0 + \frac{1}{\sqrt{2 \pi}} \int
d\omega a_\omega(x) e^{-i \omega
  \tau} + c.c. \right)$$
where $a_\omega$ are the positive frequency components of the
initial noise. The importance of this decomposition is that in
the quantum theory $a_\omega$ should identified
with the Heisenberg destruction operators.
To obtain the relation between $\epsilon_\pm$ and $a$ we linearize
eq. (\ref{E1}) to obtain
$$
A = A_0 e^{i \phi_{NL}} (1+ \epsilon/2 + i \sigma)$$ and then
compare the two solutions. The details of this comparison is given
in Appendix \ref{AppB}.

In the present work we are interested in the regime of large gain
when $gx>>1$. This implies several simplifications.  First we can
neglect $e^{-gx}$ with respect to $e^{+gx}$. Second we note that $g$
has a maximum at
\begin{equation}
\omega_{max} = \sqrt{\frac{2\gamma}{|\beta_2|}}A_0 ,
\end{equation}
hence we only need its value in the vicinity of $\omega_{max}$
\begin{equation}
g(\omega) \simeq g_{max} - |\beta_2| (\omega - \omega_{max})^2\,
\label{gomega}\end{equation} with $g_{max}=\gamma A_0^2$. Finally
all other functions can be approximated by their value at
$\omega_{max}$ (since they do not appear in exponentials, but only
as prefactors).

With these simplifications, Eqs.~\ref{AA2} lead to the following
relation between $\sigma$ and $\epsilon$
\begin{equation}
\sigma \simeq \frac{\epsilon}{2}
\end{equation}
and
the relation  between $\epsilon_\pm$ and
$a$, derived in eq. (\ref{B1}),  becomes
\begin{eqnarray}
\epsilon &\simeq&  \frac{1}{2 A_0 \sqrt{\pi}}\int_{\omega\simeq
  \omega_{max}}
\!\!\!\!\!\!\!\!\!\!\!\!\!\!
d\omega
%\nonumber\\ & &
e^{-i \omega \tau + g(\omega)x}
\left( a_\omega e^{-i \pi/4}  +
 a^*_{-\omega} e^{i\pi /4} \right)
\nonumber\\
 & & \quad + c.c.
%e^{+i \omega \tau + g(\omega)x}
%\left( a^*_{\omega} e^{i \pi /4} +
% a_{-\omega} e^{-i \pi /4} \right)\ .
\label{HB4}
\end{eqnarray}

The condition that $gx>>1$ implies that
 many photons are created in the sidebands. Thus we can neglect
quantum ordering problems and we can take $a_\omega$ to be classical
white noise with moments :
\begin{eqnarray}
&&\langle a_{\omega_1}\ldots a_{\omega_n} a^{*}_{\omega'_1}\ldots
a^*_{\omega'_m}\rangle \nonumber\\ &=&
\frac{(\hbar\omega_0)^n}{2^n}\delta_{n,m}
\sum_{\sigma}\delta(\omega_1 - \omega'_{\sigma(1)}) \ldots
\delta(\omega_{n}-\omega'_{\sigma(n)})\ ,\ \ \ \label{GG}
\end{eqnarray}
where the sum is over all permutations $\sigma$ of $\{1,\ldots
,n\}$.

In summary  we have
obtained an approximate solution of the NLSE
\begin{eqnarray}
A &\simeq& A_0 e^{i\phi_{NL}} (1 + \epsilon )^{1/2} e^{i \epsilon /2}
\label{HB5}
\end{eqnarray}
with $\epsilon$ given by eq. (\ref{HB4}) and $a_\omega$
classical white noise as described in eq. (\ref{GG}).

In order to study the power spectrum of the harmonics,
we expand eq. (\ref{HB5}) as a series in $\epsilon$ to obtain
\begin{equation}
A =
 A_0 e^{i\phi_{NL}}\sum_{n=0} c_n \epsilon^n
\label{H1},
\end{equation}
where $c_n$ are the Taylor series coefficients
\begin{eqnarray}
c_n&=&\sum_{p=0}^{n}\frac{\Gamma_{n-p}(1/2)}{(n-p)!p!}\left(\frac{i}{2}\right)^{p}\ ,\nonumber\\
\Gamma_{n}\left(\alpha\right)&=&1\times\alpha\times(\alpha-1)\ldots(\alpha-n+1)\ ,
\nonumber
\end{eqnarray}
the first few coefficients of which are:\\
 $c_0=1$,
$c_1 = \frac{1+i}{2}$, $c_2 = \frac{-1+i}{4}$ , $c_3 =
\frac{-i}{12}$ , $c_4 = \frac{-1+i}{48}$.

Recall that $\epsilon$ contains both the frequencies around
$+\omega_{max}$ and $-\omega_{max}$, see eq. (\ref{HB4}). Hence
$\epsilon^n$ contains frequencies around $n \omega_{max} , (n-2)
\omega_{max} , ..., -(n-2) \omega_{max}, -n \omega_{max}$.   Thus
each power in eq. (\ref{H1}) gives rise to a new harmonic around
frequencies $\pm n \omega_{max}$. Also there is a component around
$\omega=0$ which first arises at order $n=2$.

To compute the power
spectrum of these harmonics we take the
Fourier transform at frequency $n \omega_{max} + \delta$
of the
$n$'th order
term in eq. (\ref{H1}):
\begin{eqnarray}
&A_n(n\omega_{max}+\delta) = \frac{1}{\sqrt{2\pi}}
\int d \tau e^{i (n \omega_{max} + \delta) \tau}
c_n A_0 e^{i\phi_{NL}} \epsilon^n(\tau).&
\nonumber
%\label{An}
\end{eqnarray}
The spectral energy density at frequency $n
\omega_{max} + \delta$ is
$$P_n( n \omega_{max} + \delta) = \langle |
A_n(n\omega_{max}+\delta)
|^2\rangle \ .$$
The computation of $P_n( n \omega_{max} + \delta)$ is somewhat tedious and details are
given in Appendix \ref{AppB}. The result of the computation is:
\begin{eqnarray}
 P_n( n \omega_{max} + \delta) &=& \delta(0)
 \frac{ (\hbar\omega_{0})^{n}|c_n|^2 \sqrt{n} (n-1)! }{|A_0|^{2(n-1)}
 2^{2n-1} ( 2 \pi |\beta_2|
   x)^{(n-1)/2}}
\nonumber\\
& & \times
 \exp\left[{ 2 n g_{max} x}\right]\times \exp\left[{ - 2 |\beta_2| x \frac{\delta^2}{n}}\right]\nonumber\\
\label{MH1}\end{eqnarray}
with $n=1,2,3,\ldots$ and the first few Taylor coefficients are
$|c_0|^2 =1$, $|c_1|^2=1/2$, $|c_2|^2=1/8$, $|c_3|^2=1/144$,
$|c_4|^2=1/1152$.

The appearance of $\delta(0)$ should be interpreted as usual in this
kind of calculation as $\delta(0) = T/2\pi$ with $T$ the duration of
the light pulse. Dividing by $T$ yields the spectral power density.

For $n=1$ this describes the growth of the MI
sidebands; and for $n>1$ it describes the growth of the harmonics. As
mentioned above the first harmonic ($n=2$) also contains a component
around $\omega=0$. One can repeat the above calculation to find that
it is equal to  $4 P_2$, ie. it has exactly the same shape as the
first harmonic around $2 \omega_{max}$ except that it is centered
around $\omega = 0$ and is 4 times more intense.

\subsection{Summary of predictions and interpretation}\label{T2}

Let us first summarize the main predictions contained in eq. (\ref{MH1}); we will then discuss the interpretation and limitations of our theoretical method.
These predictions are:

\begin{enumerate}

\item
We recover the well known result that
the fundamental instability appears at frequency $\pm \omega_{max} =
\pm \sqrt{2 \gamma A_0^2 / |\beta_2|}$,
grows at a rate $2 g_{max}= 2 \gamma |A_0|^2$,
and has spectral width $\Delta \omega = \sigma_1=  1/ 2
\sqrt{|\beta_2| x}$.

\item
The power spectrum exhibits a series of Gaussian peaks centered on
frequencies $n
\omega_{max}$, $n=...,-3, -2, -1, 0, 1, 2, 3,...$ ($n=\pm1$ corresponds to the fundamental instability just mentioned).

\item The $n-1$'th harmonic appears at detuning
$\omega_n = \pm n \omega_{max}$,
grows at a rate $2g_n= 2n g_{max}= 2 n \gamma |A_0|^2$,
and has spectral width $\Delta \omega = \sigma_n=  \sqrt{n} \sigma_1
=\sqrt{n}/2 \sqrt{|\beta_2| x}$.

In addition the first harmonic has a component around $\omega=0$ which
has the same properties as the component around $2 \omega_{max}$
except that its intensity is 4 times larger.

\item

Whereas the fundamental instability can be stimulated by a classical
signal around the frequency $\omega_{max}$, the harmonics cannot be
stimulated. They are entirely determined by the initial noise around
the frequency of the fundamental instability.  We have verified this prediction experimentally by
injecting a classical signal at frequency $2\omega_{max}$ and checking
that it does not affect power spectrum of the harmonics.

\item The exact intensity of the harmonics is highly
sensitive to any classical noise present initially around the
frequency $\omega_{max}$ of the MI. Indeed the intensity of the
$n-1$'th harmonic will be multiplied by $(1/2 + n_{class})^{n}$
where $n_{class}$ is the number of classical noise photons per mode
initially present. The sensitivity thus increases with the order of
the harmonic.

\end{enumerate}

The derivation of eq. (\ref{MH1}) raises some interesting points.
First of all, the above analysis can be thought of as an expansion valid when there is no pump depletion, ie. when
the power in the modulational instability sidebands is small with
respect to the power in the pump beam.
When pump depletion can no longer be neglected our computation is no longer valid, since we supposed that the pump amplitude is constant. Furthermore in the regime where the pump gets depleted there will be a back-action of the higher order harmonics on the lower order ones, an effect we neglected (for instance we only considered the contribution of $A_n$ to $P_n$, and neglected the contributions of $A_{n'}$, $n'>n$).

The ratio $\eta$ between the energy in the fundamental sidebands at
frequency $\omega_{max}$ and $-\omega_{max}$, and the pump energy
can be explicitly computed to be
\begin{eqnarray}
\eta &=& 2\frac{\int d \delta P_1 (\omega_{max} + \delta)}{T A_0^2}
\nonumber\\
&=&\frac{\hbar \omega_0 e^{2 g_{max} x} }{4 A_0^2\sqrt{ 2 \pi
|\beta_2| x} }\nonumber
\end{eqnarray}
where the factor of $2$ takes into account that there are 2 sidebands.

The ratio of the energy in the $n-1$'th harmonic to the energy in the pump beam can then be written as
$$
2\frac{\int d \delta P_n (\omega_{max} + \delta)}{T A_0^2}
= 2 \eta^n n! |c_n|^2 $$
which shows that up to a slowly varying factor $|c_n|^2 $ the intensity in
the n-1'th harmonic is proportional to $\eta^n$. In our computation
we assumed that each successive harmonic is smaller than the
preceeding ones, ie. $P_n < P_{n-1}$. This allowed us to neglect the
back-action of higher order harmonics on lower order ones. Obviously
this corresponds to the condition $\eta <<1$.

The above analysis does not provide an exact solution of the Non
Linear Schr{\"{o}}dinger equation (\ref{HBE1}). Rather we have only
solved the linearised equations of Hasegawa and Brinkman eq.
(\ref{HBE2}). Thus our main prediction eq. (\ref{MH1}) cannot be
exact. We expect that the growth rate and spectral widths of the
harmonics are robust predictions because they depend only on the
fact that the $n-1$'th harmonic is proportional to $\epsilon^{n}$.
On the other hand the prefactor, and in particular the value of the
coefficients $|c_n|^2$ cannot be predicted correctly in the present
approach. (Indeed by using as ansatz a non linear function different
from eq. (\ref{E1}), all our results would be unchanged except the
coefficients $c_n$ which would change). Nevertheless the simple
ansatz eq. (\ref{E1}) gives surprisingly good predictions for the
intensities, see the experimental results reported in section \ref{Expt}.

Let us conclude by sketching how one could carry out a more
systematic approach to the harmonics of the MI that would predict
correctly the prefactor of eq. (\ref{MH1}). (In fact during our
first investigations of this problem we adopted this approach, but
then switched to the approach of Hasegawa and Brinkman which is much
simpler mathematically). This systematic approach is nevertheless
interesting, if only because it gives a different point of view to
the problem.

Its starting point is the standard approach of the NLSE based on the
linear equations described in Appendix \ref{AppA}. There we took an
ansatz of the form $A=(A_0 + A_1)   e^{i\phi_{NL}}$ and linearised
the equations in $A_1$ to obtain eq. (\ref{L1}). But this is only an
approximate solution. We can take into account systematic
corrections to this solution by considering the ansatz $A=(A_0 +
A_1+A_2)   e^{i\phi_{NL}}$ where $A_1$ is the solution of the
linearised equations. One then obtains for $A_2$ the equation:

\begin{equation} i \partial_x A_2 =
\frac{\beta_2}{2}\partial_\tau^2 A_2 - \gamma A_0^2 (A_2 + A_2^*)-
\gamma A_0 (2 |A_1|^2 + A_1^2). \label{LL1}
\end{equation}
This is a linear equation for $A_2$ with an independent term. Thus
the solution of the equation for $A_2$ is a solution of the
homogeneous equation plus a particular solution of the inhomogeneous
equation. The independent term is quadratic in $A_1$ and therefore
is proportional to $e^{2 g_{max}x}$ and contains frequencies around
$2 \omega_{max}$ and around $\omega=0$. Upon solving for $A_2$ one
will find that $A_2$ is largest around $2 \omega_{max}$ and around
$\omega=0$ and is proportional to $e^{2 g_{max} x}$. Thus  $A_2$
will encode the behavior of the second and zero'th harmonic.
Successive orders in perturbation theory will give rise to the
successive harmonics. We leave the detailed investigation of this
approach to future work.

\section{Experimental Results}\label{Expt}

\begin{figure}[t]
\begin{center}
\includegraphics[width=85mm]{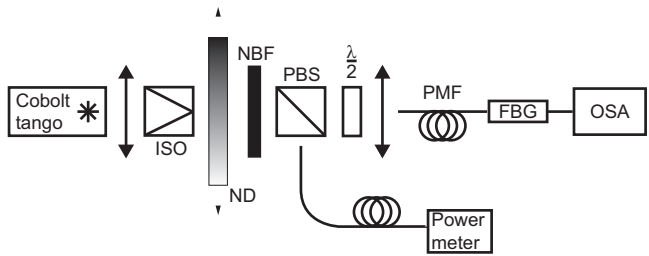}
\caption[]{Experimental setup. ISO: isolator, ND: variable neutral
  density filter, NBF: narrow band filter, PBS: polarizing beam
splitter,$\frac{\lambda}{2}$: half-wave plate, PMF: polarization
maintaining fiber, FBG: fiber Bragg grating, and OSA: optical
spectrum analyzer.} \label{montage}
\end{center}
\end{figure}

Our experimental set-up is reported in Fig.~\ref{montage}. It
consists of a Q switched laser (Cobolt Tango) that produces pulses
at 1536 nm, with a 3.55~ns full-width-at-half-maximum duration $T$
and a 2.5~kHz re\-petition rate $f$. The pump power $P_0= A_0^2$
is adjusted using a variable neutral density filter (ND). A narrow
band filter (NBF) allows a wide spectral range around the pump
wavelength to be free of noise. A polarizing beam splitter (PBS)
ensures that the pump is linearly polarized and allows to measure
the injected power proportional to the rejected beam.  We
used the Fibercore HB1250P polarizing maintaining
fiber, although the experiment could equally have been realized
with non birefringent fiber. A half-wave plate is used to ensure
that the pump polarization is aligned with a principal axis of the
fiber, whereupon polarization effects can be neglected and the
above results for scalar MI apply. The fiber parameters are
deduced from both scalar and vector modulation instabilities (see
Ref.~\cite{amansXP}). The fiber length $L$ is 51~m. The
group-velocity dispersion parameter $\beta_2$ is
$-15.27~ps^2~km^{-1}$. The Kerr nonlinearity parameter $\gamma$ is
$3.26~W^{-1}~km^{-1}$. (The beat length, which is irrelevant to
the present experiment, is 17.9~mm). Lastly, a fiber Bragg grating
(FBG) rejects the pump wavelength before the  measurement of the
spectra.  The rejection of the pump avoids detector blinding and
allows us to reach the sensibility limit of the optical spectral
analyzer (OSA).

The Narrow Band Filter (NBF) eliminates all residual
photons except those at the pump wavelength. This ensures that the
MI and harmonics indeed arise from vacuum fluctuations and not from
classical noise. We have checked that this is indeed the case in two
ways. First numerical simulations of the Stochastic Non Linear
Schr{\"{o}}dinger equation reproduce very well the observed
spectra, including the harmonics\cite{amansXP}.
Second we have studied in detail,
using a single photon detector, the MI in the regime where
relatively few photons are produced. This investigation shows that
there is in fact a small amount of noise present due to spontaneous
Raman scattering in the fiber.
But the number of Raman photons per mode
is much smaller than 1 which means that they do not affect the
spectra when the gain is large.
In summary both investigations show that the
MI process is dominated by spontaneous effects and stimulation by
classical noise is negligible.

\begin{figure}[t]
\begin{center}
\includegraphics[width=85mm]{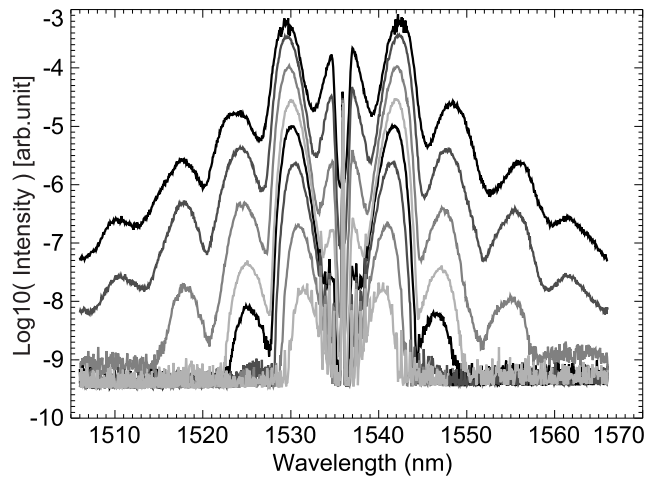}
\caption[]{Spectra of the output field for increasing peak
power $P_0$ equal to 31.6~W, 38.9~W, 45.9~W, 49.1~W, 52.2~W, 57.4~W,
59.6~W, and 62.4~W. The resolution bandwidth is 0.1~nm.
Note that the Fiber Bragg Grating has removed the pump wavelength, thereby avoiding detector blinding and allowing us to reach the sensitivity limit of the OSA.
} \label{spectra}
\end{center}
\end{figure}

A sample of the collected spectra is shown in Fig.~\ref{spectra}
for different pump powers $P_0$. We clearly observe the growth of
the MI and the appearance and growth of the harmonics as the pump
power increases. The largest pair of peaks correspond to the MI.
Then each new harmonic gives rise to a pair of peaks further and
further from the pump wavelength. In addition there is a peak
around the pump wavelength. Because this peak overlaps with the pump
it is largely rejected by the FBG and cannot be well
characterized from this first set of measurements.
In consequence we have characterized the peak around
the pump wavelength from a second set of measurements obtained
without the FBG (figure not shown). For each pump
power $P_0$, each harmonic has been fitted according to a gaussian
function $I_n(P_0) \exp\left( - \frac{(\omega -
\omega_n(P_0))^2}{2 \sigma_n^2(P_0)} \right)$ where $I_n$,
$\omega_n$, $\sigma_n$ characterize the intensity, frequency and
spectral width of the harmonic.

\begin{figure}[t]
\begin{center}
\includegraphics[width=85mm]{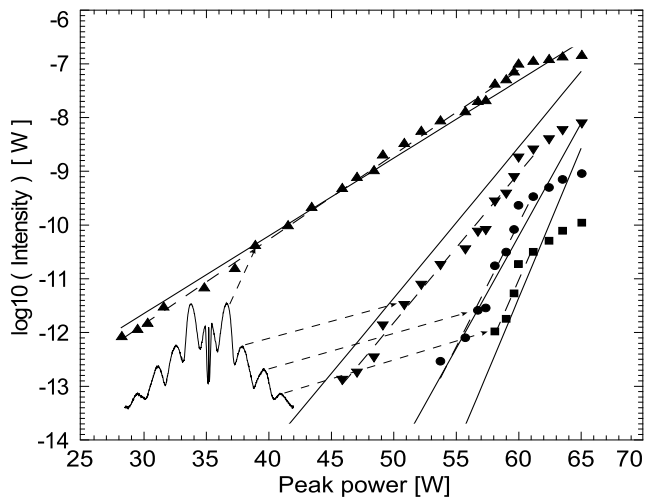}
\caption[]{Maximum intensity as a function of the peak power of the
pump. The up-triangles, down-triangles, circles, and
squares represent respectively the fundamental MI peak ($n=1$),
the first ($n=2$), second ($n=3$)
and third ($n=4$) harmonics. The dashed lines are a fit to the data
according to an exponential law. The
continuous curves are the theoretical predictions of
Eq.~\ref{MH1}.
As described in the text the experimental points were shifted by 7dB
to agree with the theoretical predictions.
For clarity, we show a spectrum.}
\label{intensity}
\end{center}
\end{figure}

The measured intensities $I_n(P_0)$ of each harmonic are compared in
Fig.~\ref{intensity} to the theoretical predictions (continuous curve).
To obtain the theoretical curves, Eq.~\ref{MH1} must be
integrated over the OSA resolution bandwidth (0.1~nm). Moreover we
must identify
$\delta(0)=\frac{T\times ST\times f}{2\pi PTS}$, where $T$ is
the pulse full width at half maximum, $ST$ is the sweep time
equal to 27.9~s, $f$ is the laser repetition rate, and $PTS$ denotes
the number of samples per spectrum equal to 1001. In order to obtain
a good overlap between theoretical curves and measured points, we had
to further shift $\delta(0)$ by 7~dB.
This discrepancy is not unreasonable
given that the above theory was based on a continuous pump, whereas
now we are dealing with a gaussian pulse, and given our
inability to
carry out
absolute measurements with the OSA.
With this shift the theoretical and experimental intensities agree
well. This shows that eq. (\ref{MH1}), including both the exponential
terms and the prefactor, correctly predicts the absolute intensities
of the harmonics. We further confirmed this agreement by fitting
the intensity of each harmonic
to an exponential law $I_n(P_0) \simeq A_{n}\exp(\alpha_{n}2\gamma
P_{0}L)$
where $A_{n}$ is an $n$~dependent constant, and $\alpha_{n}$ a growth
factor (dotted lines in Fig.~\ref{intensity}).
Theory predicts that $\alpha_n / \alpha_1 = n$. This
prediction is well verified, see
 Table~\ref{table1}.

Note that around 60~W the total energy in the sidebands becomes
comparable with the pump energy: the MI saturates and our analysis is
no longer valid.

Theory predicts that the harmonic around $\omega=0$ should
have the same growth rate, but 4 times the intensity, as the harmonic
around $2 \omega_{max}$. There is reasonable agreement concerning the
growth rate, see Table~\ref{table1}. We were unable to check the
factor of 4 as the intensity measurements of the harmonic around
$\omega =0$ were not precise enough, although it is always more
intense than the harmonic around $2\omega_{max}$, see
 Fig.~\ref{spectra}.

\begin{figure}[t]
\begin{center}
\includegraphics[width=85mm]{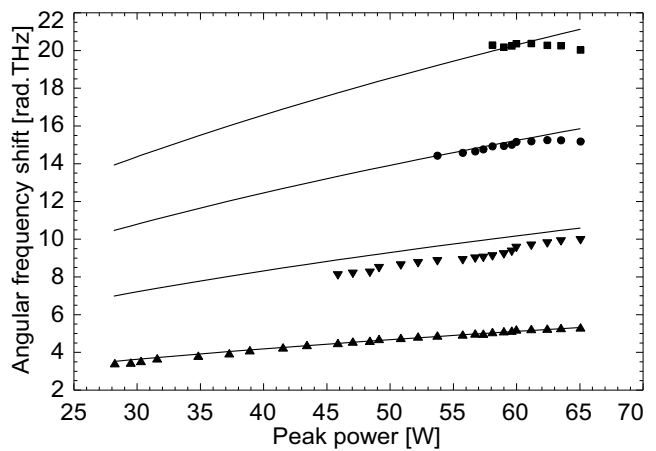}
\caption[]{Angular frequency shift as a function of the peak power
$P_0$ of the pump. The up-triangles, down-triangles, circles, and
squares represent respectively  fundamental MI peak ($n=1$), the
first ($n=2$), second ($n=3$) and third ($n=4$) harmonics. The
continuous curves correspond to the theoretical prediction $\omega_n
= n \sqrt{2 \gamma P_0/|\beta_2|}$.} \label{freqshift}
\end{center}
\end{figure}

We now turn to the angular frequency shifts $\omega_n (P_0)$. In
Fig.~\ref{freqshift} we plot the measured values and the theoretical
predictions (continuous curves). We note a very good overlap.
Moreover, from the data in Fig.~\ref{freqshift}, we have computed
the ratios between the $(n-1)^{th}$ harmonic frequency
$\omega_{n}(P_{0})$ and the fundamental frequency
$\omega_{1}(P_{0})$. The average values
$\left\langle\omega_{n}/\omega_{1}\right\rangle_{P_{0}}$ (where the
average is over the different values of $P_0$) are
reported in Table~\ref{table1}. There is good
agreement with the theoretical prediction
$\omega_{n}/{\omega_{1}}=n$.

\begin{figure}[t]
\begin{center}
\includegraphics[width=85mm]{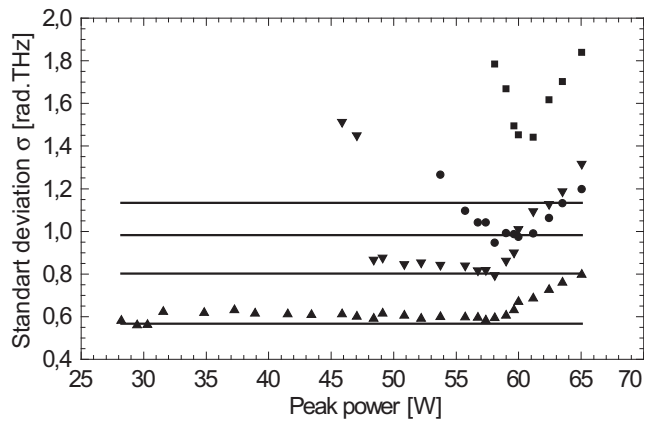}
\caption[]{The spectral width $\sigma_{n}$ of each harmonic is
shown as a function of
the peak intensity $P_0$. The up-triangles, down-triangles, circles, and
squares represent respectively the fundamental MI peak, the first, second
and third harmonics.
Straight lines correspond to the theoretical prediction:
$\sigma_n=\sqrt{n/ 4|\beta_2|L}$.
}\label{StanDev}
\end{center}
\end{figure}

Finally, the measured spectral widths $\sigma_n$ are reported in
Fig.~\ref{StanDev} as a function of $P_0$. They are compared to the
theoretical values $\sqrt{ n / 4 |\beta_2|L}$ (straight lines). The
agreement is good for the fundamental frequency ($n=1$) and the
first harmonic ($n=2$). In the case of the second ($n=3$) and third
($n=4$) harmonics it is not possible to conclude, although the
values are consistent with an increase in spectral width with the
order of the harmonic. This is because for high pump powers
($>60~W$) the MI saturates; in this regime we find that the spectral
width increases for all values of $n$. And when the peaks of the MI
harmonics have low intensity they are broadened for an unknown
reason. For the second ($n=3$) and third ($n=4$) harmonic the widths
are always affected by either of these effects. Finally the ratios
$\left\langle{\sigma_{n}^2}/{\sigma_{1}^2}\right\rangle_{P_{0}}$
(where the average is taken over those values of $P_0$ where
$\sigma_n$ is approximately constant) are reported in
Tab.~\ref{table1}. As expected, only the first harmonic leads to a
proper value.

\begin{table}[h]
    \centering
    \caption{Behavior of the harmonics. $\ast$ and $\dag$ denote
    respectively the first and the second set of measures.
The theoretical prediction is that all the quantities are equal to
    $n$ (except the center frequency of the second harmonic centered
    on $\omega = 0$).
}
\begin{tabular}{|c || *{1}{r r r|} }
\hline\hline
n (Order of&\multicolumn{3}{c|}{Measured values}\\
\cline{2-4} \rule[-8pt]{0pt}{22pt} harmonic = $n-1$)

&\large$\left\langle {\sigma_{n}^2}/ {\sigma_{1}^2}
\right\rangle_{P_{0}}$
&\large$\left\langle{\omega_{n}}/{\omega_{1}}\right\rangle_{P_{0}}$
&\large${\alpha_{n}}/{\alpha_{1}}$
     \\
\hline\hline
2 ($\omega\simeq 0$)
%&2&0    &2
&$\;2.11\pm0.03^{\dag}$&$\;0^{\dag}$&$\;2.43\pm0.15^{\dag}$\\
2 ($\omega\simeq 2 \omega_{max}$)
%&2&2    &2
&$\;2.31\pm0.02^{\dag}$&$\;1.95\pm0.01^{\dag}$&$\;1.99\pm0.12^{\dag}$\\
\hline
2 ($\omega\simeq 2 \omega_{max}$)
%&2&2    &2
&$\;2.15\pm0.07^{\ast}$&$\;1.85\pm0.01^{\ast}$&$\;1.77\pm0.06^{\ast}$\\
3
%&3&3    &3
&$\;2.21\pm0.06^{\ast}$&$\;2.95\pm0.01^{\ast}$&$\;2.92\pm0.23^{\ast}$\\
4
% &4&4    &4
&$\;4.89\pm0.15^{\ast}$&$\;3.94\pm0.03^{\ast}$&$\;3.36\pm0.61^{\ast}$\\
%\hline
%$n\in\mathrm{I\!N}^{\ast}$&$n$&$n$&$n$&&&\\
\hline\hline
\end{tabular}
\label{table1}

\end{table}

\section{Conclusion}\label{conclusion}
In summary we have studied the higher order harmonics of
Modulational Instability in the regime where the MI arises
spontaneously through amplification of vacuum fluctuations. We have
shown that in this regime there is a rich phenomenology which can be
predicted theoretically and is well verified experimentally. In
particular we have obtained predictions for the detunings, intensities, growth rates
and spectral widths of the harmonics, see our key result eq. (\ref{MH1}); and found good agreement with our experimental results based on light propagation in optical fibers. We note that the shape of the spectrum, and in particular the relative intensities of the higher order harmonics, is highly sensitive to the initial presence of classical noise, and can therefore be taken as a signature that the MI is seeded by vacuum fluctuations. It should be possible to extend our work in a number of directions, for instance to predict exactly the coefficients $|c_n|^2 $ appearing in eq. (\ref{MH1}), to study the regime where the pump starts to get depleted, and to extend it to other forms of MI, or to other kinds of instabilities.

{\bf Acknowledgments:} This research was
supported by the Interuniversity Attraction Poles
Programme - Belgium Science Policy - under grant V-18, by the Action
de Recherche Concert{\'{e}}e de la Communaut{\'{e}} Fran\c{c}aise de
Belgique, and by the Fonds Defay.

\appendix

\section{Modulational Instability}\label{AppA}

In this Appendix we recall the usual approach to MI. We first solve the linearised equations, then consider the predictions of quantum theory, and finally consider the regime of large gain. This appendix thus contains background material useful for reading the rest of the article.

\subsection{Linear Perturbation Theory}

In order to solve the NLSE eq. (\ref{NLSE}) we consider an ansatz of the form
$$A=(A_0 + A_1)  e^{i\phi_{NL}} \ .$$
where $A_1$ is a small perturbation.
We linearise
the NLSE around the continous solution to obtain the equation for $A_1$:
\begin{equation}\label{L1}
i \partial_x A_1 = \frac{\beta_2}{2}\partial_\tau^2 A_1 - \gamma A_0^2
(A_1 + A_1^*)\ .
\end{equation}
We then carry out a Fourier expansion of $A_1$:
\begin{equation}
A_1(\tau, x) = \frac{1}{\sqrt{2\pi}} \int d \omega\  a_1(x, \omega)
e^{-i \omega\tau},\label{F1}
\end{equation}
where $\omega$ is the detuning with respect to $\omega_0$.
Eq. (\ref{L1}) then becomes
\begin{equation}
\label{L2} i \partial_x a_1(x,\omega) = -\frac{\beta_2}{2}\omega^2
a_1(x,\omega) - \gamma A_0^2 \left(a_1(x,\omega) +
a_1^*(x,-\omega)\right)\ .
\end{equation}

When $\beta_2 < 0$ and when $0 \leq \omega^2 \leq \omega_c^2 = 4
\gamma A_0^2 / |\beta_2|$ this equation possesses exponentially
growing and exponentially decreasing solutions. On the other hand
when $\omega^2 > \omega^2_c$ or when $\beta_2>0$ eq. (\ref{L2})
possesses oscillating solutions. Let us consider the unstable
solutions. These are
proportional
to
$e^{\pm g x}$ where the gain is
\begin{equation}
g = \frac{|\beta_2 \omega|}{2}\left(\frac{4 \gamma
  A_0^2}{|\beta_2|} - \omega^2 \right)^{1/2} \ .
\label{L5}\end{equation}

We can reexpress the solution in terms of the initial conditions as
\begin{eqnarray}
a_1(x,\omega)&=&
a_1(0,\omega) \mu (x, \omega) + a_1^*(0,\omega) \nu
(x,\omega)\label{L3}\\
\mu (x, \omega) &=& \cosh (g x) + i \frac{(\gamma A_0^2 + \beta_2
  \omega^2 / 2)} {g} \sinh (gx)\ \ \ \\
\nu (x, \omega) &=& i \frac{\gamma A_0^2}{g} \sinh (gx),
\end{eqnarray}
where $\mu$ and $\nu$ obey the condition $|\mu|^2 - |\nu|^2 = 1$.

\subsection{Quantum Theory}

The above solution of the NLSE can also be used to derive the form
of the instability when the Modulational Instability is seeded by
vacuum fluctuations. In this case eq.~(\ref{NLSE}) should be
interpreted as the Heisenberg equation for the operators $\hat A$
and $\hat A^\dagger$. The Fourier transform eq.~(\ref{F1}) defines
the destruction operators $\hat a_1 (x, \omega)$. Their hermitian
conjugate $\hat a_1^\dagger (x,\omega)$ are the creation operators.
These operators obey the commutation relations
\begin{equation}
[\hat a_1 (x, \omega) , \hat a_1^\dagger (x, \omega')] = \hbar
(\omega_0 + \omega) \delta (\omega - \omega')\ .
\end{equation}
Equation (\ref{L3}) should then be reinterpreted as giving
 the relation between the creation and
destruction operators at distance $x$ along the fiber and at the
origin $x=0$:
\begin{equation}
\hat a_1(x,\omega)=
\hat a_1(0,\omega) \mu (x, \omega) + \hat a_1^\dagger (0,\omega) \nu
(x,\omega)\ .
\label{Q1}
\end{equation}

The initial state $|0\rangle$ contains no photons:
\begin{equation}
\hat a_1(0,\omega)|0\rangle = 0\ .
\end{equation}
From eq.~(\ref{Q1}) one can then compute the expectation values of
products of creation and destruction operators at distance $x$ along
the fiber. For instance one finds
\begin{equation}
\langle 0 | \hat a_1^\dagger (x, \omega) \hat a_1 (x,
\omega')|0\rangle = \hbar (\omega_0-\omega)|\nu(x,\omega)|^2 \delta
(\omega - \omega').\\
\label{ExpVal}
\end{equation}
From this it follows that the spectral power density at frequeny
$\omega_0+\omega$, $P(x,\omega)$, is given by
\begin{eqnarray}
 P(x,\omega) &=& \langle 0|  \hat a_1^\dagger (x, \omega)
\hat a_1 (x, \omega) |0\rangle\nonumber\\
 &=&  \hbar (\omega_0-\omega)
|\nu(x,\omega)|^2 \delta (0)\nonumber\\
&\simeq&  \hbar \omega_0 |\nu(x,\omega)|^2 \delta (0).
\label{Q2}\end{eqnarray} It is infinite because we have computed the
power when the pump beam is monochromatic and of infinite duration.
If we suppose that the pump beam lasts only for a duration $T$ then
one should -as usual in these kinds of situations- interpret this
infinity as
\begin{equation}
\delta (0) \equiv \frac{T} {2 \pi}.
\end{equation}
Then the spectral power density at position $x$ at frequency
$\omega_0 + \omega$ is $P(x,\omega)= \hbar \omega_0 |\nu|^2 T/ 2
\pi$.

Note that the quantum solution also predicts other effects such as
correlations (two mode squeezing) which will not be studied here.

\subsection{Behavior in the regime of large gain}\label{subG}

We now consider the regime where
the gain $e^{gx}$ is large. In this case we can simplify the solutions
obtained above.
First of all we note that the produced photons will be localised
around the frequency $\omega_{max}$ at which the gain is maximum:
\begin{equation}
\omega^2_{max} = \frac{2 \gamma A_0^2 }{ |\beta_2|} \label{wmax}
.\end{equation} The maximum gain is given by
\begin{equation}
g_{max} = g(\omega_{max}) = \gamma A_0^2\ .
\end{equation}

In order to describe the behavior in the vicinity of the maximum
gain we can expand the gain around $\omega_{max}$ as
\begin{equation}
g(\omega) \simeq g_{max} - \frac{g''}{2} (\omega - \omega_{max})^2,
\label{GG1}
\end{equation}
where $g''=2|\beta_2|$. In the vicinity of $\omega_{max}$ the
coefficients $\mu$ and $\nu$ simplify. They can be approximated by:
\begin{eqnarray}
\mu (x, \omega)
&\simeq&
\frac{1}{2}
\exp [ x g_{max}- x g''/2(\omega -\omega_0)^2], \nonumber\\
\nu (x, \omega) &\simeq& \frac{i}{2}\exp[x g_{max}  - x
g''/2(\omega -\omega_0)^2],\nonumber
\end{eqnarray}
where we have kept only the exponentially growing terms in $\mu$, $\nu$,
used the approximate expressions derived
in eq. (\ref{GG1}) and
dropped the $\omega$ dependence of the prefactors in $\mu$, $\nu$.

Furthermore in the regime of large gain the quantum solution
simplifies. Indeed since there are many photons in each mode one can
carry out a semiclassical treatment in which one neglects ordering
problems. Thus in this regime one can reproduce the predictions of
the quantum solution by taking the initial conditions of the
classical solution $a(0,\omega)$ to be white noise with power $\hbar
\omega_0 /2$ per mode. More precisly the $a(0,\omega)$ should be
taken to be complex delta correlated gaussian random variable
distributed according to the probability distribution
\begin{equation}
P(a(0,\omega)) = \frac{1}{2 \pi \sigma^2} e^{- |a|^2 / 2 \sigma^2},
\end{equation}
with variance $\sigma^2$ such that
\begin{equation}
\langle a(0,\omega) a^*(0,\omega')\rangle =
\frac{\hbar \omega_0 }{2}\delta(\omega -  \omega').
\end{equation}
This will correctly reproduce the quantum predictions up to
corrections proportional to $\exp[-g x]$. Indeed taking the
probability distribution to be gaussian correctly reproduces in the
regime of large gain the combinatorial factors which arise from Wick
contractions when expectation values of the products of many
creation and destruction operators are taken. Thus for instance in
this regime we have
\begin{eqnarray}
\langle a(0,\omega_1)\ldots a(0,\omega_n) a^{*}(0,\omega'_1)\ldots
a^*(0,\omega'_m)\rangle=\nonumber\\
\frac{(\hbar\omega_0)^n}{2^n}\delta_{n,m}
\sum_{\sigma}\delta(\omega_1 - \omega'_{\sigma(1)}) \ldots
\delta(\omega_{n}-\omega'_{\sigma(n)}), \label{GG2}\end{eqnarray}
where the sum is carried out over all permutations $\sigma$ of
$\{1,\ldots ,n\}$.

In the regime of large gain the spectral power density  is thus
\begin{eqnarray}
P(x,\omega) &=&
\langle a^*(x,\omega)a(x,\omega)\rangle\nonumber\\
&=& \frac{T}{2 \pi}\frac{\hbar\omega_0}{4}\exp[2 g_{max} x]\exp[  -
 g''(\omega -\omega_{max})^2 x]\ .\nonumber\\
 \label{I1}\end{eqnarray}

\section{Details of calculations}\label{AppB}

\subsection{Relation Between $\epsilon$, $\sigma$ and $a$}

Upon linearising eq. (\ref{E1}) in $\epsilon$ and $\sigma$ we obtain
\begin{eqnarray}
A &=& A_0 e^{i \phi_{NL}}  (1+ \epsilon/2 +
i \sigma)\nonumber
\end{eqnarray}
Upons inserting the forms given by eqs. (\ref{AA1}) and (\ref{AA2}) one finds
\begin{eqnarray}
A&=&A_0 e^{i\phi_{NL}} \big( 1 +
 \frac{1}{\sqrt{2\pi}} \int_0^\infty d \omega\nonumber\\
& &\quad\quad
\ \epsilon_+(\omega)(\frac{1}{2} +  \frac{ig}{|\beta_2|\omega^2})
e^{-i \omega \tau + gx}\nonumber\\
& &\quad\quad
+ \epsilon_-(\omega)(\frac{1}{2} -  \frac{ig}{|\beta_2|\omega^2})
e^{-i \omega \tau - gx}\nonumber\\
& &\quad\quad
+\epsilon^*_+(\omega)(\frac{1}{2} +  \frac{ig}{|\beta_2|\omega^2})
e^{+i \omega \tau + gx}
\nonumber\\
& &\quad\quad +\epsilon^*_-(\omega)(\frac{1}{2} -
\frac{ig}{|\beta_2|\omega^2}) e^{+i \omega \tau - gx}
\big).\nonumber\end{eqnarray}
We can then identify
\begin{eqnarray}
\epsilon_+(\omega) &=& \frac{a_1(0,\omega)}{2 A_0} \left(1 - i
  \frac{|\beta_2|\omega^2}{2 g}\right)+ \frac{a_1^*(0,\omega)}{2 A_0} \left(1 + i
  \frac{|\beta_2|\omega^2}{2 g}\right),\nonumber\\
\epsilon_-(\omega) &=& \frac{a_1(0,\omega)}{2 A_0} \left(1 + i
  \frac{|\beta_2|\omega^2}{2 g}\right)+ \frac{a_1^*(0,\omega)}{2 A_0} \left(1 - i
  \frac{|\beta_2|\omega^2}{2 g}\right). \nonumber\\
  \label{B1}\end{eqnarray}
which in the regime of large gain reduces to eq. (\ref{HB4}).

\subsection{Power Spectrum of the Harmonics}

Here we give the details of the calculations leading from eq. (\ref{H1}) to eq. (\ref{MH1}).
If we explicitise eq. (\ref{H1}) we obtain:
\begin{eqnarray}
A_{n}(x,\tau) &=& c_n A_0 e^{i\phi_{NL}} \epsilon^n \nonumber\\
&=& \frac{c_n A_0 e^{i\phi_{NL}}}{(2 \pi)^{n/2}} \int_{\omega_j
\simeq \omega_{max}}
\nonumber\\
& & \prod_{j=1}^n d \omega_j \exp\left({-i
\sum_{j=1}^n \omega_j \tau}\right) \exp\left({\sum_{j=1}^n
g(\omega_j)x}\right)
\nonumber\\
& &
\prod_{j=1}^n \left(
\frac{a_{\omega_j}}{A_0}\frac{1-i}{2} +
\frac{a^*_{-\omega_j}}{A_0}\frac{1+i}{2} \right), \label{EquDef}
\end{eqnarray}
where for conciseness we note $a_\omega = a_1(0,\omega)$. This term
contains frequencies around:\\
$-n \omega_{max},
 -(n-2)\omega_{max},\ldots,(n-2)\omega_{max},n \omega_{max}$.\\ The
 only new frequencies which appear at order $n$ are thus around $\pm
 n \omega_{max}$ (except for $n=2$ when new frequencies appear also around
$\omega=0$). It is the behavior of the $n$'th order term around
 these frequencies which we are interested in.

We thus take the Fourier component of $c_n e^{i\phi_{NL}} A_0
\epsilon^n$ at frequency $n \omega_{max} + \delta$ to obtain
\begin{eqnarray}
A_n(x, n\omega_{max}+\delta) &=&\frac{c_n A_0 e^{i\phi_{NL}}}{(2
\pi)^{(n-1)/2}} \int_{\omega_j \simeq \omega_{max}} \prod_{j=1}^n d
\omega_j
\nonumber\\
& &\delta\left(\sum_{j=1}^n \omega_j - n \omega_{max} -
\delta\right)
\nonumber\\
& &\times
 \exp\left({\sum_{j=1}^n g(\omega_j)x}\right)
\nonumber\\
& &\times
\prod_{j=1}^n \left( \frac{a_{\omega_j}}{A_0}\frac{1-i}{2} +
\frac{a^*_{-\omega_j}}{A_0}\frac{1+i}{2} \right)\ .
\nonumber\\
\label{Anomega}
\end{eqnarray}
The spectral density of power at frequency  $n \omega_{max} +
\delta$ is
\begin{eqnarray}
& &P_n( n \omega_{max} + \delta) = \langle |
A_n(n\omega_{max}+\delta)
|^2\rangle\nonumber\\
&=& \frac{|A_0|^2 |c_n|^2}{(2 \pi)^{n-1}}
\frac{(\hbar\omega_{0})^{n}}{2^n |A_0|^{2n}} \int  \prod_{j=1}^n d
\omega_j \prod_{j=1}^n d \omega'_j
\nonumber\\
& &
\delta\left(\sum_{j=1}^n \omega_j
- n \omega_{max} - \delta\right)
\delta\left(\sum_{j=1}^n \omega'_j
- n \omega_{max} - \delta\right)
\nonumber\\
& & \exp\left( {\sum_{j=1}^n ( g(\omega_j) +
g(\omega'_j)x}\right)
\nonumber\\
& &
\frac{1}{2^{n}}\sum_\sigma
2\times\delta(\omega_1 -\omega'_{\sigma(1)})\ldots
2\times\delta(\omega_n -\omega'_{\sigma(n)})\ ,
\end{eqnarray}
where the sum over $\sigma$ is a sum over all $n!$ possible
permutations.
Carrying out the integrals over $\omega'_j$ this becomes:
\begin{eqnarray}
& &P_n( n \omega_{max} + \delta) = \delta (0)
\frac{|A_0|^2 |c_n|^2}{(2 \pi)^{n-1}} \frac{(\hbar\omega_{0})^{n}n!
}{2^n |A_0|^{2n}} \int  \prod_{j=1}^n d \omega_j \nonumber\\
& &
\delta\left(\sum_{j=1}^n \omega_j - n \omega_{max} - \delta\right)
\exp\left({2 \sum_{j=1}^n g(\omega_j) x}\right) \ .
\label{B4}
\end{eqnarray}

To evaluate the remaining integral we make the change of variables
$$\omega_j = \omega_{max} + \frac{\delta}{n} + \xi_j\ .$$
In terms of these variables we have
$$ 2 \sum_{j=1}^n  g(\omega_j) = 2 n g_{max} - g''
\frac{\delta^2}{n} - g''\sum_{j=1}^n \xi_j^2$$ (since $\sum_{j=1}^n
\xi_j = 0$). We recall that $g''$ is equal to $2|\beta_{2}|$. We
thus have
\begin{eqnarray}
& &\int  \prod_{j=1}^n d \omega_j \delta\left(\sum_{j=1}^n \omega_j - n
\omega_{max} - \delta\right) \exp\left({2 \sum_{j=1}^n g(\omega_j)
x}\right) \nonumber\\
&=& \exp\left({ 2 n g_{max} x}\right) \exp({ - g''
\delta^2 x / n}) \nonumber\\
&&\times\int \prod_{j=1}^n d \xi_j \delta\left(\sum_{j=1}^n
\xi_j\right) \exp \left({ - g''\sum_{j=1}^n
\xi_j^2 x}\right)\nonumber\\
&=& \frac{1}{\sqrt{n}} \left( \frac{\sqrt{\pi}}{\sqrt{g''
x}}\right)^{n-1} \exp[{ 2 n g_{max} x}]\exp[ { - g'' x \delta^2 /
n}] \label{Int}\end{eqnarray}
where the integral over $\xi_j$ is carried out
as follows: change variables to $\zeta_j = \sum_k R_{jk} \xi_k $
where $R$ is an orthogonal matrix such that $\zeta_1 = \sum_k \xi_k
/\sqrt{n}$. Then the Jacobian of this transformation is $1$. The
integrals over $\zeta_j$ factorise into one delta function and $n-1$
gaussians, yielding eq. (\ref{Int}). Inserting this into eq. (\ref{B4})
yields eq. (\ref{MH1}).


\begin{thebibliography}{99}

\bibitem{fd} T. B. Benjamin and J. E. Freir, J. Fluid Mech. {\bf 27},
417 (1967).

\bibitem{pp} T. Taniuti and H. Washimi, Phys. Rev. Lett. {\bf 21}, 209
(1968); A. Hasegawa, Phys. Rev. Lett. {\bf 24}, 1165 (1970).

\bibitem{nlop} L. A. Ostrovskii, Sov. Phys. JETP {\bf 24}, 797 (1969)

\bibitem{HB}
A. Hasegawa and W. F. Brinkman, IEEE J. Quantum Electron. 16, 694 (1980).

\bibitem{Tai86a}
K. Tai, A. Hasegawa and A. Tomita, Phys. Rev. Lett. 56, 135 (1986).

\bibitem{Agrawal95}
G. P. Agrawal, {\em Nonlinear Fiber Optics, third ed.}, Academic
Press (San Diego), 2001.

\bibitem{bec}L. Salasnich, A. Parola, L. Reatto, Phys. Rev. Lett. {\bf
91}, 080405 (2003)

\bibitem{AK} N. N. Akhmediev and V. I. Korneev, Theor. Math. Phys. 69,
1089 (1986)

%\bibitem{LYRF} B. M. Lake, H. C. Yuen, H. Rungaldier, and
%W. E. Ferguson, J. Fluid. Mech. 83, 49 (1977)

\bibitem{VSEH} G. Van Simaeys, Ph. Emplit, M. Haelterman,
Phys. Rev. Lett. {\bf 87}, 033902 (2001)

\bibitem{amansXP}
D. Amans, E. Brainis, Ph. Emplit, M. Haelterman, S. Massar,
Optics Lett. {\bf 30}  1051 (2005)



\end{thebibliography}
\end{document}